\documentstyle[12pt]{article}
\newcommand{\be}{\begin{equation}}
\newcommand{\ee}{\end{equation}}
\newcommand{\bea}{\begin{eqnarray}}
\newcommand{\eea}{\end{eqnarray}}

\newcommand{\p}[1]{(\ref{#1})}
\newcommand{\lb}{\label}

\def\theequation{\arabic{section}.\arabic{equation}}
\begin{document}
\begin{titlepage}
\begin{flushright}
hep-th/0604111\\
April 2006
\end{flushright}
\vskip 0.9truecm

\begin{center}
{\Large\bf Master Higher-Spin Particle}

\vspace{1.5cm}
\renewcommand{\thefootnote}{\star}

{\large\bf Sergey Fedoruk}\,\footnote{On leave from Ukrainian
Engineering-Pedagogical Academy, Kharkov, Ukraine},\,\,\,{\large\bf Evgeny Ivanov}
\vspace{1cm}

   {\it Bogoliubov  Laboratory of Theoretical Physics, JINR,}\\
{\it 141980 Dubna, Moscow region, Russia} \\
\vspace{0.1cm}

{\tt fedoruk,eivanov@theor.jinr.ru}\\
\vspace{0.3cm}
\setcounter{footnote}{0}

\end{center}
\vspace{0.2cm}
\vskip 0.6truecm  \nopagebreak

\begin{abstract}
\noindent We propose a ``master'' higher-spin (HS) particle system. The particle model
relevant
to the unfolded formulation of HS theory, as well as the HS particle model with a
bosonic counterpart of supersymmetry, follow from the master model as its two
different gauges.
Quantization of the master system gives rise to a new form of the massless
HS equations in an extended space involving, besides extra spinorial coordinates, also
a complex scalar one. As solutions to these equations we recover the massless HS
multiplet
with fields of all integer and half-integer helicities, and obtain new multiplets
with a non-zero minimal helicity.
The HS multiplets are described by complex wave functions which are holomorphic in
the scalar
coordinate and carry an extra $U(1)$ charge $q\,$. The latter fully characterizes
the given multiplet by fixing the minimal helicity as $q/2$. We construct
a twistorial formulation of the master system and present the general solution
of the associate HS equations through an unconstrained twistor ``prepotential''.
\end{abstract}



\newpage

\end{titlepage}

\section{Introduction}

The latest developments in higher spin (HS) theory clearly show the importance
of (super)spaces with extra bosonic co-ordinates for
concise and suggestive formulations of this theory (see e.g.
\cite{FrVas}--\cite{BekVas}).
The simple and, at the same time, powerful device to analyze the geometric structure
of such (super)spaces is (super)particles propagating in them. For instance, the
free HS
equations in different formulations can be reproduced as a result of first
quantization of the appropriate
(super)particle \cite{BandLuk,BandLukSor,Vas}.

The model of tensorial
superparticle~\cite{BandLuk,BandLukSor}\footnote{See~\cite{B-V} for the
further related developments.}
provides a
world--line interpretation of the unfolded formulation of the HS
superfield theory~\cite{Vas}. Actually, the latter can be
equivalently given either in a hyperspace containing a
ten--dimensional bosonic subspace alongside an extra commuting Weyl
spinor, or in superspace with the Grassmann spinor coordinate
(quantization of tensorial superparticle and links of it to an
unfolded formulation were also studied
in~\cite{Vas-03,PST1,BPST,BBAST}). The ten--dimensional bosonic
subspace is parameterized by a position four-vector
$x^{\alpha\dot\beta}$ (it constitutes Minkowski subspace) and six
tensorial coordinates $z^{\alpha\beta}, \bar
z^{\dot\alpha\dot\beta}$. The crucial advantage of using this
ten--dimensional space--time is that it manifests the underlying
$OSp(1|8)$ covariance of the unfolded formulation.

There also exists a version of the unfolded formulation which makes no use of the
tensorial
coordinates~\cite{Vas,Vas-03,BBAST} (the construction of
HS conformal currents in such a formulation was a subject of recent paper~\cite{GSV}).
This version is more economic, although the $Sp(8)$ (and $OSp(1|8)$) symmetry
is hidden in its framework.
In the pure bosonic case the basic equation for the HS field $\Phi(y,\bar y,x)$
~\cite{Vas} reads
\begin{equation}\label{unfold-eq}
\left(\partial_{\alpha\dot\alpha} +i\frac{\partial}{\partial
y^{\alpha}} \frac{\partial}{\partial \bar y^{\dot\alpha}}
\right) \Phi = 0\,,
\end{equation}
where $y^{\alpha}$ is a commuting Weyl spinor,
$\bar y^{\dot\alpha}=\overline{(y^{\alpha})}$.
Solution of the unfolded equation~(\ref{unfold-eq}) can be found,
assuming the polynomial dependence of the wave function on the
commuting spinors $y^{\alpha}$, $\bar y^{\dot\alpha}$
\begin{equation}\label{wf-tens}
\Phi(x, y, \bar y) =\sum_{m=0}^{\infty}\sum_{n=0}^{\infty} y^{\alpha_1}\ldots
y^{\alpha_m} \bar y^{\dot\alpha_1}\ldots\bar y^{\dot\alpha_n}
\varphi_{\alpha_1 \ldots \alpha_m \dot\alpha_1
\ldots \dot\alpha_n}(x)\,.
\end{equation}
Independent space-time fields in the expansion of the general
field~(\ref{wf-tens}) are self--dual $\varphi_{\alpha_1 \ldots \alpha_m}$
and anti--self--dual $\varphi_{\dot\alpha_1 \ldots \dot\alpha_n}$
field strengths of all helicities (including the half--integer ones)~\cite{BBAST,GSV}.
All other fields are expressed as $x$-derivatives of these basic ones.
A classical counterpart of this formulation is the particle system with the
action~\cite{Vas}
\begin{equation}\label{act-1}
S_{1}^{HS}=\int d\tau \left(
\lambda_\alpha
\bar\lambda_{\dot\alpha}
\dot x^{\dot\alpha\alpha} + \lambda_\alpha\dot y^\alpha +
\bar\lambda_{\dot\alpha}\dot{\bar y}^{\dot\alpha} \right).
\end{equation}
The spinors $\lambda_\alpha$, $\bar\lambda_{\dot\alpha}$ are canonical
momenta for $y^{\alpha}$, $\bar y^{\dot\alpha}\,$, ``dot'' on fields denotes
the time derivative. The first class constraints
\begin{equation}\label{P-res}
P_{\alpha\dot\alpha} - \lambda_\alpha
\bar\lambda_{\dot\alpha} \approx 0
\end{equation}
after quantization reproduce the unfolded equation~(\ref{unfold-eq}).

A different model of the massless HS particle was proposed in~\cite{FedLuk}. A
distinguishing feature
of this model is its manifest covariance under the even counterpart of
$4D$ supersymmetry~\footnote{It is worth mentioning that a model with the broken
even ``supersymmetry''
was used in~\cite{FedZim,Fed} to describe a particle with fixed spin.
The even ``supersymmetry'' was also exploited in~\cite{Lecht} for the description of
spectrum of the critical
open $N=2$ string in $2+2$ dimensions.}. The action resembles
that of the usual massless $N=1$ superparticle
\begin{equation}\label{act-bsusy}
S_{2}^{HS}=\int d\tau \left( P_{\alpha\dot\alpha} \omega^{\dot\alpha\alpha} - e
P_{\alpha\dot\alpha}P^{\alpha\dot\alpha} \right),
\end{equation}
\begin{equation}\label{om}
\omega^{\dot\alpha\alpha} \equiv \dot x^{\dot\alpha\alpha} -i
{\bar\zeta}{}^{\dot\alpha}\dot{\zeta}{}^{\alpha}+i
\dot{\bar\zeta}{}^{\dot\alpha}{\zeta}^{\alpha}\,.
\end{equation}
However, the crucial difference from the superparticle case is that the Weyl spinor
$\zeta^\alpha$, $\bar\zeta^{\dot\alpha}= (\overline{\zeta^\alpha})$,
is commuting. The set of the Hamiltonian constraints in the system includes the
mass-shell constraint
$
P_{\alpha\dot\alpha}P^{\alpha\dot\alpha}\approx 0
$
and the even spinor constraints
\begin{equation}\label{cons-D}
D_\alpha\equiv \pi_\alpha
+iP_{\alpha\dot\alpha}\bar\zeta^{\dot\alpha}\approx
0\,,\qquad\qquad \bar D_{\dot\alpha}\equiv \bar\pi_{\dot\alpha}
-i\zeta^{\alpha}P_{\alpha\dot\alpha}\approx 0\,,
\end{equation}
where $\pi_\alpha$ and $\bar \pi_{\dot\alpha}$ are conjugate momenta
for $\zeta^\alpha$ and $\bar \zeta^{\dot\alpha}$.
The spinor constraints~(\ref{cons-D}) have the Poisson brackets algebra
\begin{equation}\label{alg-D}
[ D_\alpha,\bar D_{\dot\alpha}]_{{}_{P}} =2iP_{\alpha\dot\alpha}\,,
\end{equation}
which constitutes a  classical version of the even ``supersymmetry'' algebra  of
covariant derivatives with
{\it commutators} instead of anticommutators. Half of the spinor
constraints~(\ref{cons-D})
is first class whereas another half is second class. Quantization
of the system~(\ref{act-bsusy}) by Gupta--Bleuler (or analytic) method \footnote{In application
to superparticle models, this method was pioneered in \cite{GuBl}.} was performed
in~\cite{FedLuk}.
The wave function of the even ``supersymmetry'' particle obtained as a result of
this quantization procedure
is an even counterpart of chiral $N=1$ superfield
\begin{equation}\label{even-sfield}
\Psi(x_{\!\scriptscriptstyle L}, \zeta)=
\sum_{n=0}^{\infty} \zeta^{\alpha_1}\ldots \zeta^{\alpha_n}
\psi_{\alpha_1 \ldots \alpha_n}
(x_{\!\scriptscriptstyle L})
\end{equation}
depending on $x_{\!\scriptscriptstyle L}^{\dot\alpha\alpha} = x^{\dot\alpha\alpha} +i
{\bar\zeta}{}^{\dot\alpha}{\zeta}{}^{\alpha}$ and ${\zeta}{}^{\alpha}$.
Besides the chirality condition
\be
\bar D_{\dot\alpha} \Psi =0\,, \label{Chir}
\ee
this field is subjected to the equations
\begin{equation}\label{1-cl-con}
\partial_{\!\scriptscriptstyle L}^{\dot\alpha\alpha} \partial_{\alpha} \, \Psi =
0\,, \qquad
\partial_{\!\scriptscriptstyle L}^{\dot\alpha\alpha}
\partial_{{\!\scriptscriptstyle L}\,\alpha\dot\alpha}\, \Psi = 0\,,
\end{equation}
which are quantum counterparts of first class constraints.
Due to eqs.~(\ref{1-cl-con}) the fields
in the expansion of the wave function
with respect to the even spinor variables
are self--dual field strengths
of the massless particles of all helicities.
Anti--self--dual field strengths are contained
in the complex conjugated (anti-chiral) field.

The aim of this paper is to formulate and study a new (``master'') model of HS
particle which yields
both models \p{act-1} and \p{act-bsusy} upon choosing the appropriate gauges and,
after quantization,
both systems of the HS equations \p{unfold-eq} and \p{Chir}, \p{1-cl-con} as the
result of partial solving of the full set of relevant equations. One of the novel
features of this system is that,
instead of the set \p{unfold-eq}, the master model
actually  implies a modified unfolded system of equations in which the infinite
towers of higher spins are accommodated
by some holomorphic functions depending on a new scalar complex variable. These
functions are characterized by the
``external'' $U(1)$ charge number $q$ which fully specifies the corresponding
infinite-dimensional multiplet of spins and has,
as its classical counterpart, the coefficient before Fayet-Iliopoulos term in the
``master'' action.
The HS multiplets can start not only with a scalar field (as in
\p{wf-tens}), which
corresponds to the choice $q=0\,$, but also with any other self-dual field the
helicity of which is
fixed as $q/2$ by the external $U(1)$ charge.

In Sect.~2 we give a general description of our model.
The master HS particle propagates in the space
parameterized by the spinorial variables of the systems~(\ref{act-1}),
(\ref{act-bsusy}) and an
additional complex scalar. The Hamiltonian formulation of this system involves only
first
class constraints. They include unfolded constraints~(\ref{P-res}),
first class generalization of the spinor constraints~(\ref{cons-D})
and a scalar constraint which generates local $U(1)$
transformations of the twistor--like variables and complex scalar co-ordinate.

In Sect.~3 we carry out canonical quantization of the master system and
obtain the relevant set of HS equations for the wave function. This set
generalizes and involves both sets \p{unfold-eq} and \p{Chir}, \p{1-cl-con}.
Notably, the unfolded equation similar to~(\ref{unfold-eq}) proves to be a
consequence of
the quantum counterparts of the spinor constraints.
There is also a scalar $U(1)$ constraint which is an analog of the ``spin--shell''
constraint
present in the model of massless particle with fixed helicity~\cite{Fer}--\cite{Town}.
In our case the degree of homogeneity of the HS field with respect to commuting
twistor-like
spinors is not fixed due to the presence of complex scalar coordinate $\eta$ with
non-zero $U(1)$ charge. The external $U(1)$ charge $q$ of the HS wave function
in extended space is defined as
a degree of homogeneity with respect to {\it both} spinor and scalar co-ordinates.
The wave function is {\it holomorphic} in $\eta\,$.
At any fixed $q$, there is an infinite set of states in the spectrum.

At $q=0$ the HS wave function describes massless particles of all helicities
in terms of their complex self--dual field strengths (starting with a complex scalar
field).
The $q=0$ HS field
can be treated as a holomorphic ``half'' of the Vasiliev's unfolded
field~(\ref{wf-tens}).
The latter is recovered as a sum of the original HS field (holomorphic in $\eta$)
and its complex conjugate (antiholomorphic in $\eta$) which comprises
anti--self--dual field
strengths.
At $q\neq 0$ we obtain new HS multiplets having different helicity contents.
Physical states of the HS fields with $q> 0$ are massless particles with the helicities
ranging from $\frac{q}{2}$ to infinity.
When $q< 0$, the spectrum contains an infinite tower of
massless states of all positive helicities like in the $q= 0$ case and,
additionally, the
finite number of states with negative helicities described by anti--self--dual
field strengths (such HS multiplets can be named ``spin--flip'' ones).

All these HS multiplets admit an equivalent (and more economic) description
in terms of the chiral and anti-chiral fields of the bosonic ``supersymmetry''.
At $q=0$ it is just the HS field~(\ref{even-sfield}). The $q \neq 0$ multiplets are
described by the properly constrained chiral fields with external indices.

We also construct a twistor formulation of the master system.
In contrast to the twistor formulation of
the system~(\ref{act-1}), the master HS particle propagates in a
space parametrized by a unit twistor and an additional complex scalar.
We construct a coordinate twistor transform relating different
classical formulations of the master system, as well as a field twistor transform
which allows one to reconstruct the space--time HS fields by the twistorial
``prepotential''
which solves the HS equations.

In Sect.~4 we summarize our results and make some comments on the symmetries of
the master HS particle.
\setcounter{equation}{0}

\section{Master HS particle model}

\subsection{Action and constraints}

In this section we describe a new model of even HS particle effectively possessing
only first class
constraints. It plays the role of the ``master system''
both for the particle model~(\ref{act-1}) corresponding to the unfolded formulation and
for the model~(\ref{act-bsusy}) with the explicit even ``supersymmetry''.

The master system involves the variables of both systems~(\ref{act-1}) and
(\ref{act-bsusy})
and also an additional complex scalar $\eta$
($\bar\eta=(\overline{\phantom{|\!}\!\eta\!\phantom{|\!}})$).
The model is described by the following action
\begin{equation}\label{act-mast}
S^{HS}_{mast}=\int d\tau \left[\lambda_\alpha \bar\lambda_{\dot\alpha}
\omega^{\dot\alpha\alpha}
+ \lambda_\alpha\dot y^\alpha + \bar\lambda_{\dot\alpha}
\dot{\bar y}^{\dot\alpha} +i(\eta\dot{\bar\eta} -
\dot\eta\bar\eta) -
2i\bar\eta\dot{\zeta}{}^{\alpha}\lambda_\alpha +2i\eta
\bar\lambda_{\dot\alpha} \dot{\bar\zeta}{}^{\dot\alpha} - l \left({\cal N} -
c\right) \right].
\end{equation}
The action is invariant under local phase transformations of the involved complex fields except the fields $\zeta$, $\bar\zeta$, with
\be
{\cal N} \equiv i\,(y^\alpha\lambda_\alpha
-\bar\lambda_{\dot\alpha} {\bar y}^{\dot\alpha})-2\eta\bar\eta
\ee
being the $U(1)$ current, $l$ the relevant gauge field and $c$ a strength of
the $1D$ ``Fayet-Iliopoulos'' term.
The field $l$ acts as a Lagrange multiplier effecting the constraint
\begin{equation}\label{H-mast}
i\,(y^\alpha\lambda_\alpha
-\bar\lambda_{\dot\alpha} {\bar y}^{\dot\alpha})- 2\eta\bar\eta -c
\approx 0
\end{equation}
which generates, in the Hamiltonian formalism, local $U(1)$ transformations.
This constraint is the HS generalization of the ``spin-shell'' constraint
$i\,(y^\alpha\lambda_\alpha
-\bar\lambda_{\dot\alpha} {\bar y}^{\dot\alpha}) -c \approx 0$ present in
the model of massless particle with a fixed helicity~\cite{Fer}--\cite{Town}.

Hamiltonian analysis of the system~\p{act-mast}
is performed in Appendix A.
Besides the constraint~\p{H-mast}, the action~\p{act-mast} also gives rise to
the following set of the primary constraints
\begin{equation}\label{T-mast}
T_{\alpha\dot\alpha}\equiv P_{\alpha\dot\alpha} -\lambda_\alpha
\bar\lambda_{\dot\alpha}\approx 0\,,
\end{equation}
\begin{equation}\label{D-mast}
{\cal D}_\alpha\equiv D_\alpha +
2i\bar\eta\lambda_\alpha\approx 0\,,\qquad\qquad \bar{\cal D}_{\dot\alpha}\equiv
\bar{D}_{\dot\alpha} - 2i\eta\bar\lambda_{\dot\alpha}\approx 0\,,
\end{equation}
where $D_\alpha$ and ${\bar D}_{\dot\alpha}$ are defined in~(\ref{cons-D}).
Note that, by the same token as in~\cite{Fer}--\cite{Town}, the
second class constraints
\be
p_\eta +i\bar\eta \approx 0\,, \quad \bar p_{\eta} -i\eta \approx 0 \lb{Sec}
\ee
also following from \p{act-mast} can be treated in the strong sense by
introducing Dirac brackets for them.  Then the complex scalar $\eta$ and
its complex conjugate form the canonical pair
\begin{equation}\label{br-eta}
[\eta, \bar\eta ]_{{}_D}={\textstyle\frac{i}{2}}\,.
\end{equation}
The canonical brackets for another phase variables do not change
\begin{equation}\label{CPB1}
[x^{\dot\alpha\alpha} , P_{\beta\dot\beta} ]_{{}_D}
=\delta^{\alpha}_{\beta} \delta^{\dot\alpha}_{\dot\beta} \,, \quad [\zeta^{\alpha} ,
\pi_{\beta} ]_{{}_D}=\delta^{\alpha}_{\beta} \,, \quad [\bar\zeta^{\dot\alpha} ,
\bar\pi_{\dot\beta} ]_{{}_D}=\delta^{\dot\alpha}_{\dot\beta} \,,
\end{equation}
\begin{equation}\label{CPB2}
[y^{\alpha} , \lambda_{\beta} ]_{{}_D}=\delta^{\alpha}_{\beta} \,,
\quad [\bar y^{\dot\alpha} , \bar\lambda_{\dot\beta}
]_{{}_D}=\delta^{\dot\alpha}_{\dot\beta}\,.
\end{equation}
The only non-vanishing bracket of the
constraints~(\ref{H-mast})--(\ref{D-mast}) is
\begin{equation}\label{PB-con}
[{\cal D}_\alpha , \bar{\cal D}_{\dot\alpha} ]_{{}_D}=2i \, T_{\alpha\dot\alpha}\,.
\end{equation}
Therefore, all constraints~(\ref{H-mast})--(\ref{D-mast}) are effectively first
class and the quantization procedure is straightforward.

Note that one can avoid passing to Dirac brackets for the second class constraints
\p{Sec} and
quantize the whole system by Gupta-Bleuler (or analytic) method. As shown in
Appendix A,
in this approach one obtains the same set of equations on the wave function.

Let us now explain in which way the master system is gauge-equivalent to the HS
particle systems
\p{act-bsusy} and \p{act-1}.

The systems~(\ref{act-bsusy}) and (\ref{act-mast}) are (classically) equivalent
to each other in the common sector of their phase space. This sector is singled out
by choosing the
definite sign of the energy $P_0$ which is fixed due to the constraint~(\ref{T-mast}).
This equivalence becomes evident if one observes that the system~(\ref{act-mast})
can be interpreted
as the system~(\ref{act-bsusy}) in which the second class constraints are converted
into first class by
introducing new canonical pair $\eta$, $\bar\eta$.

To be more precise, we use the covariant conversion method firstly proposed
in~\cite{ES} for the case of usual superparticle.
To convert two second class constraints contained in the spinor
constraints~(\ref{cons-D}) into first class,
we introduce two additional degrees of freedom carried by the complex scalar field
$\eta$.
We also introduce a commuting Weyl spinor $\lambda_\alpha$ to ensure the Lorentz
covariance of
the new spinor constraints~(\ref{D-mast}).
The closure of the algebra~(\ref{PB-con}) of new spinor constraints
gives just the constraint~(\ref{T-mast}) resolving four-momentum in terms of the
spinor product.
This resolution is defined up to an arbitrary phase transformation of $\lambda_\alpha$
$\bar\lambda_{\dot\alpha}$. In order to ensure this $U(1)$ gauge invariance to hold
in the
full modified action, we are led to add the constraint~(\ref{H-mast}).

A heuristic argument why this equivalence should hold is that both models,
(\ref{act-bsusy}) and
(\ref{act-mast}), have the same number $n_{ph}={\bf 8}$ of the physical degrees of
freedom. The rigorous proof of
the equivalence can be achieved by reducing both systems to
the physical degrees of freedom. Namely, choosing light--cone gauge and following the
gauge-fixing procedure as in~\cite{ES}, \cite{Town}, we can show that the actions of
the systems~(\ref{act-mast})
and~(\ref{act-bsusy}) written in terms of physical variables coincide with each other
in the sector with the definite sign of energy. An exposition of this procedure is
given in Appendix B.

As for the world-line particle model~(\ref{act-1}) of the HS unfolded formulation,
it has also
eight physical degrees of freedom and so is expected to be equivalent to the
master HS particle model too. Indeed, this model also follows from the master
model~(\ref{act-mast}) with a particular gauge choice. The spinor
constraints~(\ref{D-mast}) and the gauge-fixing condition $\zeta^{\alpha} \approx 0$
together with its complex conjugate
can be used to eliminate the variables $\zeta^{\alpha}$, $\pi_\alpha$
and their complex conjugates. Then the constraint (\ref{H-mast}) can be used to gauge
away $\eta$. The phase
and the square of modulus of $\eta$ play the role of the real coordinate
and real momentum, respectively
$$
\varphi\equiv -i\ln(\eta/\bar\eta)\,, \quad \rho\equiv \eta\bar\eta\,,
\qquad [ \varphi , \rho ]_{{}_D}=1\,.
$$
The constraint~(\ref{H-mast}) is linear in $\rho$ and generates
arbitrary local transformations of $\varphi$. This constraint together with
the gauge fixing condition
\begin{equation}\label{gauge}
\chi \equiv \varphi - const \approx 0
\end{equation}
can be used to eliminate the
variables $\rho$, $\varphi$ at expense of the remaining variables $\lambda_\alpha$,
$y^\alpha$,
$\bar\lambda_{\dot\alpha}$, ${\bar y}^{\dot\alpha}$. Since the condition~(\ref{gauge})
includes only $\varphi$, the brackets for the remaining variables do not change.
As a result, we arrive at the system~(\ref{act-1}).

\subsection{Twistorial formulation of master HS particle}

Before passing to quantization, we present the twistorial formulation of the master
HS particle.
This formulation can be used to analyze symmetries of
the master HS particle model \cite{FIsoon}. The symmetry transformations, both finite-
and infinite-dimensional, prove to have a transparent form in terms of twistors.

The particle model~(\ref{act-mast}) in the twistorial formulation
is described by two Weyl spinors $\lambda_{\alpha}$ and $\bar\mu^{\dot\alpha}$
which form a twistor and by a complex scalar $\xi$. The additional spinor and scalar
are introduced by the following twistor transform
\begin{equation}\label{mu}
\mu^{\alpha} = y^{\alpha} +\bar\lambda_{\dot\beta}
(x^{\dot\beta\alpha} -i\bar\zeta^{\dot\beta}\zeta^\alpha )
- 2i\,\bar\eta\, \zeta^\alpha \, , \qquad
\bar\mu^{\dot\alpha} =\bar y^{\dot\alpha}+
(x^{\dot\alpha\beta} +i\bar\zeta^{\dot\alpha}\zeta^\beta )\lambda_{\beta} + 2i\,\eta\,
\bar\zeta^{\dot\alpha}\,,
\end{equation}
\begin{equation}\label{xi}
\xi=\eta +\zeta^\beta \lambda_\beta \, ,\qquad \bar\xi =\bar\eta
+\bar\lambda_{\dot\beta} \bar\zeta^{\dot\beta}\,.
\end{equation}
Using~(\ref{br-eta})--(\ref{CPB2}), one can check
that the newly introduced  variables are canonical ones
\begin{equation}\label{CPB-tw}
[\mu^{\alpha} , \lambda_{\beta} ]_{{}_D}=\delta^{\alpha}_{\beta} \,,
\qquad [\bar\mu^{\dot\alpha} , \bar\lambda_{\dot\beta} ]_{{}_D}
=\delta^{\dot\alpha}_{\dot\beta}\,, \qquad
[\xi, \bar\xi ]_{{}_D}={\textstyle\frac{i}{2}}\,.
\end{equation}
Up to the boundary terms, the action~(\ref{act-mast})
takes the following form in the twistorial variables
\begin{equation}\label{act-twist-2}
S^{HS-tw.}=\int d\tau \left[\lambda_\alpha\dot \mu^\alpha
+\bar\lambda_{\dot\alpha}
\dot{\bar \mu}^{\dot\alpha}+i(\xi\dot{\bar\xi} -
\dot\xi\bar\xi) - l \, \left(U - c\right)\right].
\end{equation}
Here, the $U(1)$ constraint
\begin{equation}\label{U-cons}
U - c\equiv i(\mu^\alpha\lambda_\alpha
-\bar\lambda_{\dot\alpha} {\bar\mu}^{\dot\alpha})- 2\xi\bar\xi -c
\approx 0
\end{equation}
is the condition (\ref{H-mast}) rewritten in the twistorial variables~(\ref{mu}),
(\ref{xi}).

The twistorial formulation~(\ref{act-twist-2}) of the HS particle reproduces
the twistorial HS particle which was considered in~\cite{BandLuk,BandLukSor}.
Due to the constraint~(\ref{U-cons}),
one can gauge away the variables $\xi$, $\bar\xi$. As in the previous Section,
we fix the phase of $\xi$
and eliminate the modulus of $\xi$ by the constraint~(\ref{U-cons}). The canonical
brackets for the remaining variables
$\lambda_\alpha$, $\mu^\alpha$, $\bar\lambda_{\dot\alpha}$,
$\bar\mu^{\dot\alpha}$ do not change.
As a result, we obtain the system described by
the action of ref. \cite{BandLuk,BandLukSor}
\begin{equation}\label{act-twist-1}
S_1^{HS-tw.}=\int d\tau \left(\lambda_\alpha\dot \mu^\alpha
+\bar\lambda_{\dot\alpha}
\dot{\bar \mu}^{\dot\alpha} \right).
\end{equation}
The twistor system~(\ref{act-twist-1}) can be obtained also directly
from the system~(\ref{act-1}) via the following standard twistor transform
\begin{equation}\label{inc-m}
\mu^{\alpha} = y^{\alpha} +\bar\lambda_{\dot\beta} x^{\dot\beta\alpha}
\, , \qquad \bar\mu^{\dot\alpha} =\bar y^{\dot\alpha}+
x^{\dot\alpha\beta}\lambda_{\beta} \,.
\end{equation}
This provides one more proof of the equivalence of the systems~(\ref{act-1})
and~(\ref{act-mast}).

Note that in both twistor transforms, i.e. (\ref{inc-m})
and~(\ref{mu}), (\ref{xi}), the twistor variables are defined in terms of
the variables of the ``mixed'' systems~(\ref{act-1})
and~(\ref{act-mast}) which involve both the space--time
($x$, $\zeta$) and ``twistor--like'' ($\lambda$, $y$, $\eta$) variables.
This suggests a way of computing some quantities
(for example, symmetry generators) in the ``mixed'' picture,
if we know them in the pure twistorial formulation.

For definiteness, we can choose ${\rm Arg}\,\xi=2k\pi$, $k\in
{\bf Z}$
in the gauge fixing condition for the constraint~(\ref{U-cons}). Then $\xi$ is real
and it
is expressed as
\begin{equation}\label{ksi}
\xi=\bar\xi=a(\lambda, \bar\lambda, \mu, {\bar\mu})\,,
\qquad a (\lambda, \bar\lambda, \mu, {\bar\mu})\equiv
\sqrt{{\textstyle\frac{i}{2}}(\mu^\alpha\lambda_\alpha
-\bar\lambda_{\dot\alpha} {\bar\mu}^{\dot\alpha})}
\end{equation}
due to~(\ref{U-cons}).
\setcounter{equation}{0}

\section{First-quantized theory}

\subsection{Canonical quantization of master system}

As shown in Sect. 2.1, after resolving the second class constraints \p{Sec} through
the Dirac procedure, in the system~(\ref{act-mast}) there remain only first class
constraints.
Therefore the quantization procedure is straightforward in this approach\footnote{In
Appendix A
we demonstrate that an equivalent quantization can be performed using the
Gupta-Bleuler method in which the variables $\eta$ and $\bar\eta$ are treated as
mutually conjugated in the ordinary sense and
only one of the constraints \p{Sec} is imposed on the complex wave function.}.

We use the coordinate representation
of the canonical brackets~(\ref{br-eta})--(\ref{CPB2}) in which the momentum operators
have the following realization
\begin{equation}\label{rep-3}
\hat P_{\alpha\dot\alpha}=-i\, \partial/\partial
x^{\dot\alpha\alpha}\equiv -i\, \partial_{\alpha\dot\alpha}\,,
\qquad \hat\pi_{\alpha}=-i \,\partial/\partial \zeta^{\alpha}
\equiv -i \partial_{\alpha} \,, \qquad \hat{\bar\pi}_{\dot\alpha}=-i
\,\partial/\partial \bar\zeta^{\dot\alpha} \equiv -i
\bar\partial_{\dot\alpha}\,,
\end{equation}
\begin{equation}\label{rep-4}
\hat\lambda_{\alpha}=-i \,\partial/\partial y^{\alpha} \,,
\qquad \hat{\bar\lambda}_{\dot\alpha}=-i \, \partial/\partial
\bar y ^{\dot\alpha} \,, \qquad \hat{\bar\eta} ={\textstyle\frac{1}{2}}\,
\partial/\partial \eta\,.
\end{equation}
The constraints~(\ref{H-mast})--(\ref{D-mast}) become the set of equations for the
wave function
$F^{(q)} (x, \zeta, \bar\zeta, y, \bar y, \eta)$
\begin{equation}\label{T-eq}
\left(\partial_{\alpha\dot\beta} +i\frac{\partial}{\partial
y^{\alpha}}
\frac{\partial}{\partial \bar y ^{\dot\beta}} \right) F^{(q)} = 0\,,
\end{equation}
\begin{equation}\label{D-eq}
\left(D_{\alpha}+ \frac{\partial}{\partial
\eta} \frac{\partial}{\partial
y^{\alpha}} \right) F^{(q)} = 0\,,
\end{equation}
\begin{equation}\label{bD-eq}
\left(\bar D_{\dot\alpha}- 2\eta \frac{\partial}{\partial
\bar y^{\dot\alpha}} \right)\, F^{(q)} = 0\,,
\end{equation}
\begin{equation}\label{H-eq}
\left(y^{\alpha}\frac{\partial}{\partial
y^{\alpha}} - \bar y^{\dot\alpha}\frac{\partial}{\partial
\bar y^{\dot\alpha}} - \eta\frac{\partial}{\partial
\eta} \right) F^{(q)} = q\,F^{(q)}\,.
\end{equation}
Here, the operators
$$
D_\alpha= -i (\partial_{\alpha}
+i\partial_{\alpha\dot\alpha} \bar\zeta^{\dot\alpha} )
\,,\qquad {\bar D}_{\dot\alpha}= -i
(\bar\partial_{\dot\alpha}
-i\zeta^{\alpha}\partial_{\alpha\dot\alpha})
$$
are quantum counterparts of the ``covariant momenta''~(\ref{cons-D}).

The external $U(1)$ charge $q$ defined by~(\ref{H-eq}) is the quantum counterpart of
the ``classical'' constant $c$ present in the constraint~(\ref{H-mast}), now with
the ordering ambiguities taken into account. Eq.~(\ref{H-eq}) implies $U(1)$
covariance of the wave function
\begin{equation}\label{u1-inv}
F^{(q)} (x, \zeta, \bar\zeta, e^{i\varphi} y,
e^{-i\varphi} \bar y, e^{-i\varphi} \eta) =
e^{qi\varphi} F^{(q)} (x, \zeta, \bar\zeta, y, \bar y, \eta)\,.
\end{equation}
Requiring $F^{(q)}$ to be single-valued restricts $q$ to the integer values. As
follows from
\p{u1-inv}, the $U(1)$ charges of the coordinates $y_\alpha$ and
$\bar{y}_{\dot\alpha}$ are opposite while $\eta$ has the same charge as
$\bar{y}_{\dot\alpha}\,$. The rest of coordinates are
neutral. It is important to note that this $U(1)$ covariance just implies that any
monomial
of the charged coordinates in the series expansion of the wave function $F^{(q)}$
has the same fixed charge $q$ and it does not entail any $U(1)$ transformation of
the coefficient fields.
In this respect $q$ resembles the ``harmonic $U(1)$ charge'' of the
harmonic superspace approach \cite{GIOS,GIKOS} and the operator in \p{H-eq} is an
analog
of the $U(1)$ charge-counting operator $D^0$ in this approach.
As we shall see, $\frac{q}{2}$ always coincides with the lowest helicity in the
infinite tower
of helicities comprised by the wave function $F^{(q)}\,$.

Before presenting the solution of eqs.~\p{T-eq}--\p{H-eq}, let us mention one more
notable feature of this system. Though it involves the Vasiliev-type unfolded
vector equation~(\ref{T-eq}), the latter follows from the spinorial eqs.~(\ref{D-eq})
and~(\ref{bD-eq}) as their integrability condition\footnote{One can say that \p{D-eq}
and \p{bD-eq} play the role of bosonic ``square root'' of \p{T-eq}.}.
So the basic independent equations of the system \p{T-eq}--\p{H-eq} are the bosonic
spinorial equations (\ref{D-eq}), (\ref{bD-eq}) and the $U(1)$ condition \p{H-eq}.

Let us now solve eqs.~(\ref{D-eq})--(\ref{H-eq}). One can do this in several
equivalent ways.

It is most convenient to work in the variables $\zeta^{\alpha}$,
$\bar\zeta^{\dot\alpha}$,
$y^{\alpha}$, $\eta$ and
\begin{equation}\label{x,by-l}
x_{\!\scriptscriptstyle L}^{\dot\alpha\alpha} =
x^{\dot\alpha\alpha} + i\,
\bar\zeta^{\dot\alpha}\zeta^{\alpha}\,, \qquad
\bar y_{\!\scriptscriptstyle L}^{\dot\alpha} =
\bar y^{\dot\alpha} + 2i\,
\eta\,\bar\zeta^{\dot\alpha}\,.
\end{equation}
In these new variables eqs.~(\ref{D-eq})--(\ref{H-eq}) take the form
\begin{equation}\label{D-eq-1}
\left[ -i \left(\partial_{\alpha} +i \frac{\partial}{\partial \eta}
\frac{\partial}{\partial y^{\alpha}}\right) + 2
\bar\zeta^{\dot\alpha} \left(\partial_{{\!\scriptscriptstyle
L}\,\alpha\dot\alpha} +i\frac{\partial}{\partial y^{\alpha}}
\frac{\partial}{\partial \bar y_{\!\scriptscriptstyle
L}^{\dot\alpha}} \right) \right] F^{(q)} = 0\,,
\end{equation}
\begin{equation}\label{bD-eq-1}
\bar\partial_{\dot\alpha} F^{(q)} =0\,,
\end{equation}
\begin{equation}\label{H-eq-1}
\left(y^{\alpha}\frac{\partial}{\partial
y^{\alpha}} - \bar y_{\!\scriptscriptstyle L}^{\dot\alpha}\frac{\partial}{\partial
\bar y_{\!\scriptscriptstyle L}^{\dot\alpha}} - \eta\frac{\partial}{\partial
\eta} \right) F^{(q)} = q \, F^{(q)}\,.
\end{equation}

Eq.~(\ref{bD-eq-1}) is the bosonic chirality condition stating that $F^{(q)}$ does not
depend on $\bar\zeta^{\dot\alpha}$ in the new variables,
\begin{equation}\label{Phi-left}
F^{(q)} = F^{(q)} (x_{\!\scriptscriptstyle L}, \zeta, y, \bar
y_{\!\scriptscriptstyle L}, \eta)\,.
\end{equation}
Then eq.~(\ref{D-eq-1}) amounts to two equations
\begin{equation}\label{D-eq-1a}
\left(\partial_{\alpha} +i \frac{\partial}{\partial
\eta} \frac{\partial}{\partial
y^{\alpha}}\right) F^{(q)} = 0
\end{equation}
and
\begin{equation}\label{T-eq-1}
\left(\partial_{{\!\scriptscriptstyle L}\,\alpha\dot\alpha}
+i\frac{\partial}{\partial y^{\alpha}} \frac{\partial}{\partial \bar
y_{\!\scriptscriptstyle L}^{\dot\alpha}} \right) F^{(q)} = 0\,.
\end{equation}
The important corollaries of eqs.~(\ref{D-eq-1a}) and (\ref{T-eq-1}) are
the following equations
\begin{equation}\label{Su-eq}
\partial_{\!\scriptscriptstyle L}^{\dot\alpha\alpha} \partial_{\alpha} \, F^{(q)} =
0\,, \qquad
\partial_{\!\scriptscriptstyle L}^{\dot\alpha\alpha}
\partial_{{\!\scriptscriptstyle L}\,\alpha\dot\alpha}\, F^{(q)} = 0\,,
\end{equation}
where we exploited the property $\frac{\partial}{\partial y^{\alpha}}
\frac{\partial}{\partial y_{\alpha}} \equiv 0$.

As the next step, we extract some further consequences of eq.~(\ref{D-eq-1a}).
By analogy with~(\ref{even-sfield}) we assume the polynomial dependence
of the field (\ref{Phi-left}) on $\zeta^\alpha$
\begin{equation}\label{Phi-z-exp}
F^{(q)}(x_{\!\scriptscriptstyle L}, \zeta, y,
\bar y_{\!\scriptscriptstyle L}, \eta)=
\sum_{n=0}^{\infty} \zeta^{\alpha_1}\ldots \zeta^{\alpha_n}
\Phi^{(q)}_{\alpha_1 \ldots \alpha_n}
(x_{\!\scriptscriptstyle L}, y,
\bar y_{\!\scriptscriptstyle L}, \eta) \,.
\end{equation}
Eq.~(\ref{D-eq-1a}) expresses all the coefficients in this expansion
as derivatives of the first
coefficient $\Phi^{(q)}(x_{\!\scriptscriptstyle L},
 y, \bar y_{\!\scriptscriptstyle L}, \eta)$
\begin{equation}\label{sol-Phi-z}
\Phi^{(q)}_{\alpha_1 \ldots \alpha_n}  =
\frac{(-i)^n}{n!} \,\frac{\partial^n}{\partial
\eta^n} \,\,\frac{\partial}{\partial
y^{\alpha_1}}\cdots \frac{\partial}{\partial
y^{\alpha_n}}\,\, \Phi^{(q)}(x_{\!\scriptscriptstyle L},
y, \bar y_{\!\scriptscriptstyle L},  \eta)\,.
\end{equation}
Thus we have reduced the initial system of equations to the two
equations on the function $\Phi^{(q)}$ \be
\mbox{(a)}\;\;\left(\partial_{{\!\scriptscriptstyle
L}\,\alpha\dot\alpha} +i\frac{\partial}{\partial y^{\alpha}}
\frac{\partial}{\partial \bar y_{\!\scriptscriptstyle
L}^{\dot\alpha}} \right) \Phi^{(q)} = 0\,, \quad
\mbox{(b)}\;\;\left(y^{\alpha}\frac{\partial}{\partial y^{\alpha}} -
\bar y_{\!\scriptscriptstyle L}^{\dot\alpha}\frac{\partial}{\partial
\bar y_{\!\scriptscriptstyle L}^{\dot\alpha}} -
\eta\frac{\partial}{\partial \eta} \right) \Phi^{(q)} = q \,
\Phi^{(q)}\,. \lb{RedSys} \ee

Like in the previous cases, we assume that $\Phi^{(q)}
(x_{\!\scriptscriptstyle L}, y,
\bar y_{\!\scriptscriptstyle L}, \eta)$ has
a non-singular polynomial expansion
over the additional coordinates. Then eq.~(\ref{RedSys}b) implies
\be
\Phi^{(q)}(x_{\!\scriptscriptstyle L}, y, \bar y_{\!\scriptscriptstyle L}, \eta) = \sum_{k=0}^\infty \eta^k \varphi^{(q+k)}(x_{\!\scriptscriptstyle L},
y, \bar y_{\!\scriptscriptstyle L})\,, \lb{quplusk}
\ee
and
\be
\left(y^{\alpha}\frac{\partial}{\partial
y^{\alpha}} - \bar y_{\!\scriptscriptstyle L}^{\dot\alpha}\frac{\partial}{\partial
\bar y_{\!\scriptscriptstyle L}^{\dot\alpha}}\right) \varphi^{(q + k)} =
(q + k)\, \varphi^{(q +k)} \lb{RedCond}
\ee
(recall that $\eta $ carries the negative unit $U(1)$ charge). The reduced $U(1)$
condition
\p{RedCond} fixes the $y, \bar y$ dependence of the functions $\varphi^{(q+k)}$ as
\begin{equation}\label{field-gen}
\varphi^{(q+k)}(x_{\!\scriptscriptstyle L}, y,
\bar y_{\!\scriptscriptstyle L})= \!
\left\{ \!
\begin{array}{lr}
\sum\limits_{n=0}^{\infty} y^{\alpha_{1}}\ldots y^{\alpha_{q+k+n}}\bar
y_{\!\scriptscriptstyle L}^{\dot\beta_1}
\ldots \bar y_{\!\scriptscriptstyle L}^{\dot\beta_n}\,
\phi_{\alpha_{1} \ldots \alpha_{q+k+n}
\dot\beta_1 \ldots
\dot\beta_n}(x_{\!\scriptscriptstyle L})\,, &
\,\,\,\,\,(q+k)\!\geq 0, \\
\sum_{n=0}\limits^{\infty} y^{\alpha_{1}}\ldots y^{\alpha_{n}}\bar
y_{\!\scriptscriptstyle L}^{\dot\beta_1}
\ldots \bar y_{\!\scriptscriptstyle L}^{\dot\beta_{|q+k|+n}}\,
\phi_{\alpha_{1} \ldots \alpha_{n}
\dot\beta_1 \ldots
\dot\beta_{|q+k|+n}}(x_{\!\scriptscriptstyle L})\,, &
\,\,\,\,\, (q+k)\!< 0.\\
\end{array}
\right.
\end{equation}

It remains to reveal the restrictions imposed on the fields $\varphi^{(q+k)}$
by the remaining unfolded equation (\ref{RedSys}a). We shall separately consider the
cases $q =0, q>0$ and $q<0\,$.

\subsubsection{q = 0}

It is easy to see that in this case eq.~(\ref{RedSys}a) expresses all the fields
$\phi_{\alpha_1 \ldots \alpha_{k+n}
\dot\beta_1 \ldots \dot\beta_n}$ with $n > 0$ in $\varphi^{(k)}$
as $x$-derivatives of the lowest component,
the self--dual field $\phi_{\alpha_1 \ldots \alpha_k}\,$.
The latter field satisfies Dirac and Klein--Gordon equations
\begin{equation}\label{eqs-D,KG}
\partial^{\dot\beta\alpha_1}\phi_{\alpha_1 \ldots
\alpha_k}  =0\,, \qquad \Box\,\phi =0
\end{equation}
also in consequence of the same eq.~(\ref{RedSys}a).

Thus the space of physical states of the model is spanned by
the complex self--dual field strengths
$\phi_{\alpha_1 \ldots \alpha_k}$,
$k=0,1,2, \ldots\,$, of the massless particles of all integer and half-integer
helicities.

It is to the point here to make comparison with the Vasiliev
approach~\cite{Vas,GSV}.
The field $\Phi^{(0)}$ and its conjugate $\bar\Phi^{(0)}(x_{\!\scriptscriptstyle R}, y_{\!\scriptscriptstyle R}, \bar y, \bar\eta)$
(with $x_{\!\scriptscriptstyle R} = \overline{(x_{\!\scriptscriptstyle L})}$,
$y_{\!\scriptscriptstyle R} =
\overline{(\bar y_{\!\scriptscriptstyle L})}$)
encompass the same set of fields as the real field~(\ref{wf-tens}), the only
difference
being the presence of {\it complex} scalar field in our case compared to a real such
field
in (\ref{wf-tens}).
Indeed, the conjugated HS field
$\bar\Phi^{(0)}(x_{\!\scriptscriptstyle R}, y_{\!\scriptscriptstyle R},
\bar y, \bar\eta)$ describes anti-self-dual fields which are complex conjugates of
the self-dual fields $\phi_{\alpha_1 \ldots \alpha_k}\,$. So no any doubling of
higher spins
occurs in our case, like in the original unfolded formulation.
Formally, the HS field~(\ref{wf-tens}) can be recovered as the sum $\Phi
=\Phi^{(0)} +\bar\Phi^{(0)}$ at the ``point'' $\eta =1$.

Thus we conclude that the higher spin multiplet
comprised by the field~(\ref{wf-tens})
is a particular case of the multiplets arising after quantization of the master HS particle,
corresponding to the choice $q=0\,$ for the external $U(1)$ charge.

\subsubsection{q $>$ 0}

Like in the $q=0$ case, for the generic positive $q$ eq.~(\ref{RedSys}a) expresses the fields
$\phi_{\alpha_1 \ldots \alpha_{q+k+n}
\dot\beta_1 \ldots \dot\beta_n}$ with $n > 0$ in~\p{field-gen}
in terms of the $ \partial_{\alpha\dot\alpha}$-derivatives
of the self-dual fields $\phi_{\alpha_1 \ldots \alpha_{q+k}}$. Also,
the same eq.~(\ref{T-eq-1}) yields Dirac equations for the independent fields
\begin{equation}\label{eqs-D}
\partial^{\dot\beta\alpha_1}\phi_{\alpha_1 \ldots
\alpha_{q+k}}  =0 \,, \qquad k=0,1,2,\ldots\,.
\end{equation}

Thus the space of physical states of the model is spanned by the self--dual field
strengths of the massless particles with helicities $\frac{q}{2},\frac{q}{2}+\frac{1}{2},\frac{q}{2}+1,
\ldots\,$. We observe that the scalar
field is absent in the spectrum for non-zero positive $q\,$. Once again, the
relevant HS multiplet is fully characterized by the value of $q\,$.

\subsubsection{q $<$ 0}

For the negative values of $q$
the expansion~(\ref{quplusk}) can be conveniently rewritten as
\begin{eqnarray}\label{field-1-neg}
\Phi^{(q)}(x_{\!\scriptscriptstyle L}, y,
\bar y_{\!\scriptscriptstyle L}, \eta) &=&
\eta^{|q|}\tilde\Phi^{(0)}(x_{\!\scriptscriptstyle L}, y,
\bar y_{\!\scriptscriptstyle L}, \eta) +\\
&& + \sum_{k=1}^{|q|}
\sum_{m=0}^{\infty} \eta^{|q|-k}\, y^{\alpha_1}\ldots y^{\alpha_{m}}
\bar y_{\!\scriptscriptstyle L}^{\dot\beta_1}
\ldots \bar y_{\!\scriptscriptstyle L}^{\dot\beta_{m+k}}\,
\phi_{\alpha_1 \ldots \alpha_{m}
\dot\beta_1 \ldots
\dot\beta_{m+k}}(x_{\!\scriptscriptstyle L})\,.
\nonumber
\end{eqnarray}
Thus in this case we deal with the HS field $\tilde\Phi^{(0)}$ having the same
helicity contents
as the $q=0$ multiplet and an additional term involving the space--time fields
$\phi_{\alpha_1 \ldots \alpha_{m}\dot\beta_1 \ldots \dot\beta_{m+k}}\,$. The set of
independent fields in this expansion consists of self--dual fields
$\phi_{\alpha_1 \ldots \alpha_k}$, $k=0,1,2, \ldots$ present in the HS field $\Phi^{(0)}$
and anti--self--dual fields $\phi_{\dot\beta_1 \ldots \dot\beta_{k}}$, $k=1,\ldots, |q|$.
Eq.~(\ref{RedSys}a) expresses all other space--time fields as
space--time derivatives of these basic ones. Eq.~(\ref{RedSys}a) also implies
the Dirac and Klein--Gordon equations for the basic fields.

Thus, for $q< 0$ physical fields in the spectrum describe
massless particles with all positive helicities starting from the zero one,
and also a finite number of massless states with negative helicities $-\frac{1}{2},
-1, \ldots,
-\frac{|q|}{2}$. This HS multiplet could be naturally called ``helicity-flip''
multiplet.
Note that, being considered together with its conjugate, this multiplet reveals a
partial doubling
of fields with a given helicity, the phenomenon which is absent in the previous two
cases.

We would like to point out that the multiplets listed above were
derived under the assumption that the $\eta$-expansions of the wave
functions are regular at the origin and so are given by power
series over $\eta $. Alternatively, we could consider the wave
functions which are regular at $\infty $ and so are represented by
series over $1/\eta\,$. In this case we would obtain basically the same
set of states (up to the interchange of positive and negative helicities).
If one gives up the regularity assumption, it is impossible to avoid the doubling of
helicities for any value of $q\,$.

\subsection{Alternative description}

Here we establish the link with the HS description suggested by the quantization of
the
particle with bosonic ``supersymmetry''~\cite{FedLuk}.

We start with the case $q=0$ and consider at first the expansion
of the field $F^{(q)}$ defined in~(\ref{Phi-left}) with respect to the coordinates
$y$, $\bar y_{\!\scriptscriptstyle L}$ and $\eta$
\begin{equation}\label{Phi-left-1}
F^{(0)} (x_{\!\scriptscriptstyle L}, \zeta, y, \bar y_{\!\scriptscriptstyle L},
\eta) = \sum_{k=0}^{\infty}
\sum_{n=0}^{\infty} \eta^{k}\, y^{\alpha_1}\ldots y^{\alpha_{k+n}}
\bar y_{\!\scriptscriptstyle L}^{\dot\alpha_1}
\ldots \bar y_{\!\scriptscriptstyle L}^{\dot\alpha_n}\,
\Psi_{\alpha_1 \ldots \alpha_{k+n}
\dot\alpha_1 \ldots
\dot\alpha_n}(x_{\!\scriptscriptstyle L}, \zeta) \,,
\end{equation}
where we have already taken care of eq.~(\ref{H-eq-1}). It remains
to take into account eqs.~(\ref{D-eq-1a}) and~(\ref{T-eq-1}). They express all
the coefficient fields in~\p{Phi-left-1} as derivatives of the lowest coefficient
$\Psi(x_{\!\scriptscriptstyle L}, \zeta)\,$:
\begin{equation}\label{sol-Phi-y1}
\Psi_{\alpha_1 \ldots \alpha_k \alpha_{k+1} \ldots \alpha_{k +n}
\dot\alpha_1 \ldots \dot\alpha_n}  =
{\textstyle\frac{i^{k+n}}{k!(n+k)!n!}}
\,\,\partial_{\alpha_1}\ldots \partial_{\alpha_k} \,\,
\partial_{{\!\scriptscriptstyle L}\alpha_{k+1}
\dot\alpha_1}\ldots \partial_{{\!\scriptscriptstyle L}\alpha_{k +n}
\dot\alpha_n}\,\, \Psi\,.
\end{equation}
The field $\Psi (x_{\!\scriptscriptstyle L}, \zeta)$ satisfies
eqs.~(\ref{Su-eq}) and it is exactly the chiral HS
field~(\ref{even-sfield}) of the paper~\cite{FedLuk}. It comprises
the same irreducible HS $q=0$ multiplet as $\Phi^{(0)}
(x_{\!\scriptscriptstyle L}, y, \bar{y}_{\!\scriptscriptstyle L},
\eta)\,$. The possibility to describe the same multiplet in two
equivalent ways is just the quantum version of equivalence between
the HS particles \p{act-1} and \p{act-bsusy} which both follow from
the master HS particle. In a sense, the description by the field
$\Psi(x_{\!\scriptscriptstyle L},\zeta)$ is more economical since
this quantity collects, in its $\zeta^\alpha$ expansion, just the
independent space-time self-dual fields (anti-self-dual fields are
contained in the complex--conjugated function $\bar\Psi
(x_{\!\scriptscriptstyle R}, \bar\zeta)\,$).

Let us also consider the alternative description of the $q > 0$ HS multiplets.
The corresponding counterpart of the $q=0$ expansion is
\begin{equation}\label{field-q-pos}
F^{(q)}(x_{\!\scriptscriptstyle L}, \zeta, y,
\bar y_{\!\scriptscriptstyle L}, \eta)= \sum_{k=0}^{\infty}
\sum_{n=0}^{\infty} \eta^{k}\, y^{\alpha_1}\ldots y^{\alpha_{q+k+n}}
\bar y_{\!\scriptscriptstyle L}^{\dot\beta_1}
\ldots \bar y_{\!\scriptscriptstyle L}^{\dot\beta_n}\,
\Psi_{\alpha_1 \ldots \alpha_{q+k+n}
\dot\beta_1 \ldots
\dot\beta_n}(x_{\!\scriptscriptstyle L}, \zeta)\,,
\end{equation}
where we have already taken account of the $U(1)$ condition (\ref{H-eq-1}).
Then  eqs.~(\ref{D-eq-1a}) and~(\ref{T-eq-1}) lead to the following expressions for
the coefficients in~\p{field-q-pos}
\begin{equation}\label{sol-Phi-q}
\Psi_{\alpha_1 \ldots \alpha_q \alpha_{q+1} \ldots \alpha_{q+k+n} \dot\beta_1 \ldots
\dot\beta_n}  =
{\textstyle\frac{i^{k+n}q!}{k!n!(q+n+k)!}} \,\partial_{\alpha_{q+1}}\ldots
\partial_{\alpha_{q+k}}
\,\partial_{{\!\scriptscriptstyle L}\alpha_{q+k+1}
\dot\beta_1}\ldots \partial_{{\!\scriptscriptstyle L}\alpha_{q+k+n}
\dot\beta_n}\, \Psi_{\alpha_1 \ldots \alpha_q}.
\end{equation}
Thus all component fields are expressed in terms of the even counterpart
of $N=1$ chiral field with external indices
$\Psi_{\alpha_1 \ldots \alpha_q} (x_{\!\scriptscriptstyle L}, \zeta)\,$.

Also as a consequence of eqs.~(\ref{D-eq-1a}) and~(\ref{T-eq-1}), this field
proves
to be subjected to eqs.~(\ref{Su-eq}) and also to the equations
\begin{equation}\label{Dir-q}
\partial^{\alpha_1}\Psi_{\alpha_1 \ldots \alpha_q} =0\,, \qquad
\partial_{\!\scriptscriptstyle L}^{\dot\alpha\alpha_1}
\Psi_{\alpha_1 \ldots \alpha_q} =0  \,.
\end{equation}
Expanding the field $\Psi_{\alpha_1 \ldots \alpha_q}$ in powers of $\zeta_\alpha$
\begin{equation}\label{Phi-exp-q}
\Psi_{\alpha_1 \ldots \alpha_q}
(x_{\!\scriptscriptstyle L}, \zeta)
= \sum_{n=0}^{\infty} \zeta^{\beta_1}\ldots \zeta^{\beta_n}
\psi_{\alpha_1 \ldots \alpha_q \beta_1 \ldots \beta_n}
(x_{\!\scriptscriptstyle L})\,,
\end{equation}
we observe that all component fields in this expansion
are totally symmetric in the spinor indices due to the first equation
in~(\ref{Dir-q}). As a consequence of eqs.~(\ref{Su-eq}) and
the second equation in~(\ref{Dir-q}), all component fields satisfy Dirac equation.
Therefore, the HS field \p{Phi-exp-q} describes the same physical spectrum
as in eq.~(\ref{eqs-D}). Note that the bosonic chiral ``superfields'' with
extra indices were not considered in~\cite{FedLuk}. It is the master HS particle
framework which suggests the importance of considering such superfields
as providing an alternative description of the $q>0$ HS multiplets.

For the $q<0$ case, a formulation in terms of the fields depending on
$x_{\!\scriptscriptstyle L}$ and $\zeta$ can be also easily constructed. It operates
with
the HS field~(\ref{even-sfield}) comprising the infinite set of positive helicities
and few extra anti--self--dual fields $\phi_{\dot\beta_1 \ldots \dot\beta_{k}} (x_{\!\scriptscriptstyle L})\;
(k = 1, \ldots, |q|)\,$.

\subsection{Quantization in twistor formulation: twistor transform for HS fields}

In the remainder of this Section we show that
quantization of the considered model in the twistorial formulation~(\ref{act-twist-2})
gives rise to the same results.

In the ``twistorial representation'' \footnote{The spinors $\lambda_\alpha$
and $\bar\mu^{\dot\alpha}$ together form a unit twistor; they both are diagonal in
the ``twistorial representation''~\cite{PenMac}.}, when $\lambda_\alpha$,
$\bar\mu^{\dot\alpha}$ and $\xi$ are diagonal and
$$
\hat{\mu}^\alpha= i\partial/\partial \lambda_\alpha\,, \quad
\hat{\bar\lambda}_{\dot\alpha}= -i
\partial/\partial \bar\mu^{\dot\alpha}\,, \quad
\hat{\bar\xi}= {\textstyle\frac{1}{2}}\,
\partial/\partial \xi\,,
$$
the twistorial wave function
\begin{equation}\label{wf-tw}
G^{(q-2)}(\lambda, \bar\mu, \xi)
\end{equation}
satisfies the quantum counterpart of the constraints~(\ref{U-cons})
\begin{equation}\label{H-eq-tw}
\left(-\lambda_{\alpha}\frac{\partial}{\partial
\lambda_{\alpha}} - \bar \mu^{\dot\alpha}\frac{\partial}{\partial
\bar \mu^{\dot\alpha}} - \xi\frac{\partial}{\partial
\xi} \right) G^{(q-2)} = (q-2)\,G^{(q-2)}\,.
\end{equation}

The wave function in the space--time description
is derived by analogy with the standard twistor approach~\cite{PenMac}.
One substitutes the incidence conditions~(\ref{mu})
and~(\ref{xi}) for the variables $\bar\mu^{\dot\alpha}$
and $\xi$ in the wave function~(\ref{wf-tw}) and performs a Fourier-type integral
transformation from $\lambda_\alpha$ to its canonically conjugated variable $y_\alpha$
\begin{equation}\label{int-tr}
F^{(q)} (x_{\!\scriptscriptstyle L}, \zeta, y,
\bar y_{\!\scriptscriptstyle L}, \eta) =
\int d^2\lambda\,  e^{i y^\alpha \lambda_\alpha}
G^{(q-2)}(\lambda_\alpha, \bar y_{\!\scriptscriptstyle L}^{\dot\alpha}+
x_{\!\scriptscriptstyle L}^{\dot\alpha\beta} \lambda_{\beta} ,
\eta +\zeta^\beta \lambda_\beta)\,.
\end{equation}
Note that the variables $x_{\!\scriptscriptstyle L}$ and
$\bar y_{\!\scriptscriptstyle L}$ defined in~(\ref{x,by-l})
already appeared in the twistor transform~(\ref{mu}) for $\bar\mu$. The
integrand in~(\ref{int-tr}) includes the Fourier exponent in contrast to the Penrose
integral
transform~\cite{PenMac}.

Using the particular dependence of the twistorial field $G^{(q-2)}\,$ on the
involved co-ordinates,
it is easy to check that the field $F^{(q)}$ defined by~(\ref{int-tr}) automatically
satisfies eqs.~(\ref{D-eq-1a}) and (\ref{T-eq-1}).
Also, homogeneity condition~(\ref{H-eq-1}) follows
from eq.~(\ref{H-eq-tw}). Thus,
the twistorial formulation solves
eqs.~(\ref{H-eq-1}), (\ref{D-eq-1a}) and (\ref{T-eq-1}) for the field~(\ref{Phi-left})
in terms of the unconstrained ``prepotential'' field~(\ref{wf-tw}).
\setcounter{equation}{0}

\section{Summary and outlook}

In this paper we have constructed a ``master'' model of HS particle.
It possesses several notable features which are summarized below
together with some of their corollaries:

\begin{itemize}

\item Both the unfolded HS particle and the HS particle with the even
``supersymmetry''
can be reproduced from the master model upon some gauge fixings.

\item Hamiltonian formulation of the master system involves
only first class constraints that, in particular, makes it possible
to directly pass to the first-quantized model.

\item The twistor formulation of the master system is a HS generalization
of the well--known twistor formulation of massless particles with fixed
helicities.

\item After quantization, the set of equations for the HS wave function in an extended space
includes the Vasiliev-type unfolded equation. But the latter turns out to be a consequence of some
basic spinorial equations related to the even ``supersymmetry'' of the model.

\item The HS fields respect a local $U(1)$ symmetry which is similar
to the $U(1)$ covariance of the harmonic approach \cite{GIOS,GIKOS}. Crucial for maintaining
this covariance is the holomorphic dependence of the HS wave function on the new complex
bosonic coordinate $\eta\,$.

\item Depending on their
external $U(1)$ charge $q$, the HS fields in the extended space accommodate
different HS multiplets of ordinary $4D$ fields. The all-helicity HS
multiplet of the unfolded formulation (with a complex scalar field) is recovered
as the $q=0$ multiplet and its conjugate, and also some new HS multiplets with
$q \neq 0$ emerge. For $q>0$
they are spanned by the self--dual field strengths of growing positive helicities,
starting from $\frac{q}{2}$. The $q<0$ multiplets show up an interesting
``spin--flip''
feature: they include self-dual fields of all positive helicities, as well as a
finite number of
anti--self--dual fields with negative helicities. The complementary helicities are
accommodated by complex conjugated wave functions.

\item The unfolded formulation of HS fields and their formulation
with explicit bosonic ``supersymmetry'' are equivalent to each other as they
correspond to different ways of solving the same master system of
HS equations. For the new HS multiplets (with $q\neq 0$) this equivalence implies,
besides the appropriate unfolded formulation, also the description
with the explicit bosonic ``supersymmetry''.

\item The master HS equations in the twistorial formulation can be solved in terms
of unconstrained twistor prepotential field.

\end{itemize}
In the rest of this Section we announce some results of studying symmetries of
the master HS particle model and the new HS equations associated with it. More details
will be given in \cite{FIsoon}.

Since the master model encompasses both the unfolded HS particle~(\ref{act-1})
and the HS particle with explicit even ``supersymmetry''~(\ref{act-bsusy}), it should
respect the symmetries inherent to both these systems.
Moreover, the equivalence of these models implies that they possess the same
full sets of symmetries, though realized in different ways.
Some symmetries can be explicit in the first model, but hidden in the second
one, and vice versa.

Let us illustrate this on the example of
the invariance of the system~(\ref{act-bsusy})
under the even ``supersymmetry'' translations~\cite{FedLuk}--\cite{Fed}
\begin{equation}\label{susy}
\delta x^{\dot\alpha\alpha} = i(\bar\epsilon^{\dot\alpha}\zeta^\alpha
-\bar\zeta^{\dot\alpha}\epsilon^\alpha) \, ,\quad \delta
\zeta^\alpha = \epsilon^\alpha \, ,\quad \delta
\bar\zeta^{\dot\alpha}= \bar\epsilon^{\dot\alpha}\,,
\end{equation}
where $\epsilon^\alpha$ is a commuting Weyl spinor.
Master system~(\ref{act-mast}) is invariant
under the transformations~(\ref{susy}) too.
The variables $\lambda_\alpha$, $y^\alpha$, $\eta$ and their
complex conjugates are inert under the even ``supersymmetry''
translations.

The symmetry~(\ref{susy}) is hidden in the
particle model~(\ref{act-1}). However, it becomes
manifest in the twistorial formulations~(\ref{act-twist-2}),
(\ref{act-twist-1}) \footnote{The advantage
of using twistors~\cite{PenMac,Fer} for the analysis of
symmetries of relativistic particles was demonstrated in~\cite{T}
(for recent developments in the twistor approach
see~\cite{B} and references therein).}.
In the twistorial formulation~(\ref{act-twist-2})
of the master system, the transformations~(\ref{susy}) act as
\begin{equation}\label{var-mu,xi}
\delta\lambda_{\alpha}=\epsilon^\alpha\lambda_\alpha\,,
\qquad
\delta \mu^{\alpha}= -2i\,\bar\xi \epsilon^{\alpha} \quad \mbox{and c. c.}\,.
\end{equation}
    They are generated by the following charges
\begin{equation}\label{R}
R_{\alpha}=-2i\, \bar\xi\lambda_{\alpha}\,, \qquad \bar
R_{\dot\alpha}=2i\,\xi\bar\lambda_{\dot\alpha}\,.
\end{equation}
The twistorial model~(\ref{act-twist-1}) is obtained from the model~(\ref{act-twist-2})
via the gauge fixing. Therefore, bosonic ``supersymmetry'' translations in
the model~(\ref{act-twist-1}) should be accompanied by the appropriate
compensating gauge transformation in order to preserve the gauge-fixing condition.
As a result we obtain the following (nonlinear) even ``supersymmetry''
transformations of the variables in the model~(\ref{act-twist-1})
\begin{equation}\label{var-l-R}
\delta\lambda_{\alpha}=-{\textstyle\frac{1}{2}}\,a^{-1}
(\epsilon\lambda -\bar\lambda
{\bar\epsilon})\lambda_{\alpha}\,, \qquad
\delta \mu^{\alpha}={\textstyle\frac{1}{2}}\,a^{-1}
(\epsilon\lambda -\bar\lambda
{\bar\epsilon})\mu^{\alpha}-2ia\epsilon^{\alpha}\,,
\end{equation}
where $a(\lambda, \bar\lambda, \mu, {\bar\mu})$
was defined in~(\ref{ksi}). It is straightforward to check that these transformations
leave invariant the action~(\ref{act-twist-1}).
Correspondingly, the bosonic ``supersymmetry'' generators in the
model~(\ref{act-twist-1}) are
\begin{equation}\label{R1}
R_{\alpha}=-2i a \lambda_{\alpha}\,, \qquad \bar
R_{\dot\alpha}=2i a \bar\lambda_{\dot\alpha}\,.
\end{equation}

The model~(\ref{act-twist-1}) has also $Sp(8)$ symmetry.
It is generated by the full set of bilinear products of the
spinors $\lambda_{\alpha}$,
$\bar\lambda_{\dot\alpha}$, $\mu^{\alpha}$ and
$\bar\mu^{\dot\alpha}$.
The closure of $Sp(8)$ with bosonic ``supersymmetry'' generated
by~(\ref{R1}), as well as with an infinite sequence of phase transformations
generated by arbitrary powers of $a(\lambda, \bar\lambda, \mu, {\bar\mu})$,
constitutes the full infinite-dimensional symmetry
of the model~(\ref{act-twist-1}). This infinite-dimensional invariance
can be interpreted as simplectic diffeomorphisms~\cite{SS}--\cite{IMT}.
A similar interpretation of an infinite-dimensional symmetry in the case of the
twistor
particle with a fixed helicity was given in~\cite{T}.

As follows from the form of the generators~(\ref{R1}), on the HS
fields~(\ref{wf-tens})
depending on the coordinates $x$, $y$, $\bar y$  the bosonic ``supersymmetry''
transformations are
realized by highly non--linear (and presumably non--local) transformations.
Remarkably, on the fields $\Phi^{(q)}$ defined in~(\ref{quplusk}) and depending on
the extra
bosonic co-ordinate $\eta$, the even ``supersymmetry'' generators have the simple form
\begin{equation}\label{R-gen}
R_{\alpha}=-\,\frac{\partial}{\partial\eta}
\, \frac{\partial}{\partial y^\alpha}\,, \qquad \bar
R_{\dot\alpha}=2 \eta
\frac{\partial}{\partial \bar y^{\dot\alpha}}\,.
\end{equation}
They close just on the $x$-translations with the generator
$\sim \partial_{\alpha\dot\alpha}$, taking into account the unfolded equation
(\ref{RedSys}a). This example clearly shows the important role the co-ordinate
$\eta$ plays in revealing symmetries of the master system of HS equations. It is
worthwhile to
notice that the transformations with the generators \p{R-gen} mix up fields with
different helicities from different coefficients $\varphi^{(q+k)}$ in \p{quplusk}.

As for further directions of the study, constructing supersymmetric extensions of the
proposed model (with standard ``odd'' supersymmetries) is one of the urgent problems.
Supersymmetric $N=1$ HS theories constructed on the basis of theories
with bosonic ``supersymmetry'' are expected to be the simplest HS theories respecting
the fundamental notion of chirality which underlies the geometric approach
to $N=1$ supergravity~\cite{OS}. The existence of a chiral limit
seems to be crucial for any satisfactory HS superfield theory including
HS generalizations of $N=1$ supergravity~\cite{BPST,IvLuk}.

Another issue to be further explored is related to the presence of commuting
spinorial variables
in most of versions of the HS field theory. In our opinion, the question of how to
ensure the correct statistics of the component fields in the expansions of HS field
over even spinors (see e.g. the power expansion~(\ref{wf-tens}) and other analogical
expansions)
is still open (see e.g. some comments in~\cite{FedLuk}). We expect
that the consideration of the HS models having both even and odd supersymmetries
could provide a clue to resolving this problem.

Finally, it would be interesting to investigate possible implications of the
additional scalar
coordinate $\eta$ (which plays a crucial role in the master system) in the geometry
of interacting HS
fields and in the problem of constructing the relevant actions. Also, it is
desirable to see
how the master system could be extended to incorporate the tensorial coordinates.

\section*{Acknowledgments}
We thank J.A.~de Azc\'arraga, I.~Bandos, J.~Lukierski, D.~Sorokin and M.~Vasiliev for interest
in the work and useful remarks. E.I. thanks the
Physical Department of the National Technical University in Athens for the kind
hospitality at the final stage
of this work. A partial support from the RFBR grant 06-02-16684 and a grant of
Heisenberg-Landau
program is acknowledged.

\renewcommand\theequation{A.\arabic{equation}} \setcounter{equation}0
\section*{Appendix A: Analysis of the constraints and Gupta--\break
\hspace*{4.2cm} Bleuler quantization}

In this Appendix we carry out more detailed Hamiltonian analysis of
the master system~(\ref{act-mast}) and show that it can be
equivalently quantized by the Gupta--Bleuler (or ``analytic'')
method~\cite{GuBl}.

Like in any system with one time derivative on fields, the expressions for all
momenta give the constraints. Namely, we obtain the primary
constraints~(\ref{T-mast}), (\ref{D-mast}) and the
constraints
\begin{equation}\label{g-mast}
g \equiv p_\eta +i\bar\eta \approx 0\,,\qquad \bar g \equiv \bar p_{\eta} -i\eta
\approx 0\,,
\end{equation}
\begin{equation}\label{pl-mast}
p_l \approx 0\,,
\end{equation}
where $p_\eta$, $\bar p_{\eta}$ and $p_l$ are the conjugate momenta for $\eta$,
$\bar \eta$ and $l\,$, respectively. We at once treat the variable $\lambda_\alpha$
and $\bar\lambda_{\dot\alpha}$ as momenta for $y^\alpha$ and $\bar
y^{\dot\alpha}\,$. The total Hamiltonian
\begin{equation}\label{Hamil}
H_T= l ({\cal N} -c)+ \Lambda^{\alpha\dot\alpha} T_{\alpha\dot\alpha} +
\Lambda^{\alpha}
{\cal D}_{\alpha} + \bar{\cal D}_{\dot\alpha} \bar\Lambda^{\dot\alpha} + v g + \bar
v \bar g + \gamma p_l
\end{equation}
is a sum of the canonical Hamiltonian $H_0= l ({\cal N}-c)$ and a linear combination of
the constraints with Lagrange multipliers $\Lambda^{\alpha\dot\alpha}$, $\ldots$\,,
$\gamma$.

The requirement of preservation of the constraint~(\ref{pl-mast}) produces the
constraint~(\ref{H-mast}) as the secondary constraint. Due to the
constraint~(\ref{pl-mast}) the variable $l$ has the pure gauge character. It plays
the role of Lagrange multiplier for the constraint~(\ref{H-mast}).

The preservation of the constraints~(\ref{g-mast}) gives rise to the elimination of
a part of
Lagrange multipliers in terms of the remaining ones
$$
v=-\Lambda^\alpha \lambda_\alpha - i l\eta\,, \quad \bar v=-
\bar\lambda_{\dot\alpha} \bar\Lambda^{\dot\alpha}
+ i l\bar\eta\,.
$$
Inserting these expressions in the Hamiltonian~(\ref{Hamil}), we obtain
\begin{equation}\label{Ham1}
H_T= l (\tilde{\cal N}-c)+ \Lambda^{\alpha\dot\alpha} T_{\alpha\dot\alpha}
+ \Lambda^{\alpha} \tilde{\cal D}_{\alpha}
+ \bar{\tilde{\cal D}}_{\dot\alpha} \bar\Lambda^{\dot\alpha}\,,
\end{equation}
where the new constraints are
\begin{equation}\label{H-mast1}
\tilde{\cal N}-c \equiv {\cal N} - i (\eta g -\bar\eta \bar g) -c
= i(y^\alpha\lambda_\alpha
-\bar\lambda_{\dot\alpha} {\bar y}^{\dot\alpha})
- i (\eta p_\eta -\bar\eta \bar p_\eta) -c
\approx 0\,,
\end{equation}
\begin{equation}\label{D-mast1}
\tilde{\cal D}_\alpha\equiv {\cal D}_\alpha - g \lambda_\alpha
= D_\alpha -(p_\eta-i\bar\eta)\lambda_\alpha\approx 0\,,
\end{equation}
\begin{equation}\label{bD-mast1}
\bar{\tilde{\cal D}}_{\dot\alpha} \equiv \bar{\cal D}_{\dot\alpha} -\bar g
\bar\lambda_{\dot\alpha} = {\bar D}_{\dot\alpha} - (\bar p_\eta+ i\eta)
\bar\lambda_{\dot\alpha}\approx 0\,.
\end{equation}

All constraints present in the Hamiltonian~(\ref{Ham1}) are first class. The only
non-vanishing
Poisson bracket of them is
\begin{equation}\label{PB-con1}
[\tilde{\cal D}_\alpha , \bar{\tilde{\cal D}}_{\dot\alpha} ]_{{}_P}=2i
T_{\alpha\dot\alpha}\,.
\end{equation}

The constraints~(\ref{g-mast}) commute with the constraints~(\ref{T-mast}),
(\ref{H-mast1})--(\ref{bD-mast1}) but have a non-vanishing Poisson bracket between
themselves
\begin{equation}\label{PB-g}
[g , \bar g ]_{{}_P}=2i\,.
\end{equation}
So they are second class. We can introduce the Dirac brackets for them. Then,
the constraints~(\ref{H-mast1})--(\ref{bD-mast1}) become~(\ref{H-mast}),
(\ref{D-mast})
and the variables $\eta$, $\bar\eta$ satisfy commutation relation~(\ref{br-eta}).
The quantization goes as in Sect. 3.

Here we outline the alternative Gupta--Bleuler approach to the treatment of
the constraints~(\ref{g-mast}). When quantizing in this approach, the wave function
is assumed to depend on both $\eta$ and $\bar\eta$ as the coordinates. Their momenta
have the standard realization
$$
\hat p_\eta=-i \,\partial/\partial \eta \,,
\qquad \hat{\bar p}_{\eta}=-i \, \partial/\partial
\bar \eta\,.
$$
We impose on the wave function
$$
\tilde F^{(q)}(x, \zeta, \bar\zeta, y, \bar y, \eta, \bar \eta )
$$
all first class constraints~(\ref{T-mast}), ~(\ref{H-mast1})--(\ref{bD-mast1}) and
half of the second class ones. For definiteness, we take the constraint $\bar g$.
Thus, the equations on the wave
function are
\begin{equation}\label{g-eq1}
-i\left( \frac{\partial}{\partial
\bar \eta} + \eta \right) \tilde F^{(q)} = 0\,,
\end{equation}
\begin{equation}\label{T-eq1}
\left(\partial_{\alpha\dot\beta} +i\frac{\partial}{\partial
y^{\alpha}}
\frac{\partial}{\partial \bar y ^{\dot\beta}} \right)
\tilde F^{(q)} = 0\,,
\end{equation}
\begin{equation}\label{D-eq1}
\left[D_{\alpha}+ \frac{\partial}{\partial
y^{\alpha}} \left( \frac{\partial}{\partial
\eta} +\bar\eta \right) \right] \tilde F^{(q)} = 0\,,\qquad
\left[\bar D_{\dot\alpha}+ \frac{\partial}{\partial
\bar y^{\dot\alpha}} \left(\frac{\partial}{\partial
\bar\eta} -\eta\right) \right] \tilde F^{(q)} = 0\,,
\end{equation}
\begin{equation}\label{H-eq1}
\left(y^{\alpha}\frac{\partial}{\partial
y^{\alpha}} - \bar y^{\dot\alpha}\frac{\partial}{\partial
\bar y^{\dot\alpha}} - \eta\frac{\partial}{\partial
\eta} + \bar\eta\frac{\partial}{\partial
\bar\eta} \right) \tilde F^{(q)} = q\,\tilde F^{(q)}\,.
\end{equation}

Eq.~(\ref{g-eq1}) is solved by
\begin{equation}\label{tPhi-s}
\tilde F^{(q)}(x, \zeta, \bar\zeta, y, \bar y, \eta, \bar \eta ) = e^{-\eta\bar\eta}
F^{(q)}(x, \zeta, \bar\zeta, y, \bar y, \eta )\,,
\end{equation}
where the reduced wave function $F^{(q)}$ does not depend on $\bar\eta$.
Inserting the solution~(\ref{tPhi-s}) in the remaining
eqs~(\ref{T-eq1})--(\ref{H-eq1}), we
find that $F^{(q)}$ obeys just eqs.~(\ref{T-eq})--(\ref{H-eq}) which were
obtained in Sect. 3.

\renewcommand\theequation{B.\arabic{equation}} \setcounter{equation}0
\section*{Appendix B: Equivalence of the master model
and \break
\hspace*{4.27cm} the model with bosonic ``supersymmetry''}

In this Appendix we prove the classical equivalence of
the master system~(\ref{act-mast}) and the HS
particle~(\ref{act-bsusy}) with bosonic ``supersymmetry''.
The proof is based on reducing both systems to physical degrees of freedom.

We shall use the light--cone notations
\begin{equation}\label{lc-x}
x^+ \equiv x^{\dot 1 1} ={\textstyle\frac{x^0+x^3}{\sqrt{2}}}\,, \qquad x^- \equiv
x^{\dot 2 2} ={\textstyle\frac{x^0-x^3}{\sqrt{2}}}\,, \qquad x^{(tr)} \equiv x^{\dot
2 1} ={\textstyle\frac{x^1+ ix^2}{\sqrt{2}}}\,,
\end{equation}
\begin{equation}\label{lc-p}
P_+ \equiv P_{1\dot 1} ={\textstyle\frac{P_0+P_3}{\sqrt{2}}}\,, \qquad P_- \equiv
P_{2\dot 2} ={\textstyle\frac{P_0-P_3}{\sqrt{2}}}\,, \qquad P^{(tr)} \equiv P_{2\dot
1} ={\textstyle\frac{P_1+ iP_2}{\sqrt{2}}}\,.
\end{equation}
\vspace{0.3cm}

\noindent{\it The system~(\ref{act-bsusy}) in physical variables.}
In the notation~(\ref{lc-x}), (\ref{lc-p}) the constraints of the
system~(\ref{act-bsusy})
take the following form
\begin{equation}\label{mass-1}
P_+ P_- -|P^{(tr)}|^2 \approx 0\,,
\end{equation}
\begin{equation}\label{D-1}
D_1= \pi_1
+iP_+ \bar\zeta^{\dot 1} +i\bar P^{(tr)} \bar\zeta^{\dot 2} \approx
0\,,\qquad\qquad \bar D_{\dot 1}= \bar\pi_{\dot 1}
-i P_+ \zeta^{1} -i \bar P^{(tr)}\zeta^{2}\approx 0\,,
\end{equation}
\begin{equation}\label{D-2}
D_2= \pi_2
+i P^{(tr)} \bar\zeta^{\dot 1} +i P_- \bar\zeta^{\dot 2} \approx
0\,,\qquad \bar D_{\dot 2}= \bar\pi_{\dot 2}
-i P^{(tr)} \zeta^{1} -i P_- \zeta^{2}\approx 0\,.
\end{equation}
We choose the light--cone gauge (it is assumed that $P_+ \neq 0$)
\begin{equation}\label{lc-ga-1}
u^-\equiv x^- -\tau \approx 0\,,
\end{equation}
\begin{equation}\label{lc-ga-2}
\zeta^{2} \approx
0\,,\qquad \bar\zeta^{\dot 2}\approx 0\,.
\end{equation}

The constraints~(\ref{D-2}) and the gauge fixing conditions~(\ref{lc-ga-2})
eliminate phase variables $\zeta^{2}$, $\pi_{2}$ and their complex conjugates
without affecting the commutation relations for the
remaining variables. But the gauge~(\ref{lc-ga-1}) involves the proper time.
To correctly account for it, we make a canonical transformation from the system with
the coordinate
$x^-$ to the system with the coordinate $u^-$ defined by the
relation~(\ref{lc-ga-1}). The generating function of this canonical transformation
is given by
\begin{equation}\label{gen-f}
F = P_-^{(u)} (x^- -\tau) + \cdots\,,
\end{equation}
where dots stand for the identical transformation for other variables. This
generating function
produces the
relation~(\ref{lc-ga-1}) and yields $P_-^{(u)}= P_-$. Thus, the
conditions~(\ref{lc-ga-1}) and~(\ref{mass-1}) eliminate $u^-$ and imply
\begin{equation}\label{p-}
P_- = \frac{1}{P_+}\,|P^{(tr)}|^2 \,.
\end{equation}
We also should take into account that the canonical transformation depends on the
proper time and, therefore, generates the new Hamiltonian (initial Hamiltonian is
equal to zero)
\begin{equation}\label{Ham}
H= \frac{\partial F}{\partial \tau}= -P_- = -\frac{1}{P_+}\,|P^{(tr)}|^2 \,.
\end{equation}
As a result, in physical variables the system~(\ref{act-bsusy}) is described by the
action
\begin{equation}\label{act-1-ph}
S^{ph}=\int d\tau \left( P_+ \dot x^+ + \bar P^{(tr)} \dot x^{(tr)} + P^{(tr)} \dot
{\bar x}^{(tr)} \pm i(\dot{\bar\eta}\eta -\bar\eta\dot\eta) \mp
{\textstyle\frac{1}{|P_+|}}|P^{(tr)}|^2 \right),
\end{equation}
where we introduced $\eta = \sqrt{|P_+|}\zeta^1$. In the action~(\ref{act-1-ph})
the upper (lower) sign is chosen for $P_+$ positive (negative).
\vspace{0.3cm}

\noindent{\it The system~(\ref{act-mast}) in physical variables.}
The first class constraint~(\ref{D-mast}) and the gauge-fixing condition for it
\begin{equation}\label{z-fix}
\zeta^\alpha \approx 0\,,\qquad \bar\zeta^{\dot\alpha}\approx 0
\end{equation}
eliminate the variables $\zeta^\alpha$, $\pi_\alpha$ and their complex conjugates
without affecting the Poisson brackets for other variables. The remaining
constraints~(\ref{H-mast}) and (\ref{T-mast}) take the form
\begin{equation}\label{H-ph}
{\cal N}-c \equiv p^{(\phi)}-2\eta\bar\eta -c
\approx 0\,,
\end{equation}
\begin{equation}\label{P+}
P_+ - (r_1)^2\approx 0\,,
\end{equation}
\begin{equation}\label{P-}
P_- - (r_2)^2\approx 0\,,
\end{equation}
\begin{equation}\label{P-tr}
P^{(tr)} - r_1 r_2 e^{-2i\psi}\approx 0\,,
\qquad
\bar P^{(tr)} - r_1 r_2 e^{2i\psi} \approx 0\,,
\end{equation}
where the variables $r_1$, $r_2$, $\phi$, $\psi$ are defined by
\begin{equation}\label{l-1,2}
\lambda_1 = r_1 e^{i(\phi+\psi)}\,,
\qquad
\lambda_2 = r_2 e^{i(\phi-\psi)}.
\end{equation}
The generating function of the canonical transformation from
the variables $\lambda_\alpha$, $y^\alpha$ and their complex conjugates to
the variables $r_1$, $r_2$, $\phi$, $\psi$ and conjugate momenta of the latter
is given by
\begin{equation}\label{gen-f2}
F=y^1 r_1 e^{i(\phi+\psi)} +y^2 r_2 e^{i(\phi-\psi)} + {\it c. c.}
\end{equation}
{}From this expression we obtain the relation $p^{(\phi)}= i(y\lambda - \bar\lambda
\bar y)$
used in the constraint~(\ref{H-ph}).

The constraints~(\ref{H-ph}), (\ref{P+}), (\ref{P-tr}) together with the
gauge-fixing conditions
for them,
\begin{equation}\label{g-fix}
\phi \approx 0\,,\qquad p^{(r)}_1 \approx 0\,,\qquad p^{(r)}_2 \approx 0\,,\qquad
p^{(\psi)}
\approx 0\,,
\end{equation}
completely eliminate the degrees of freedom present in the spinors $\lambda_\alpha$,
$y^\alpha$
and their complex conjugates.

The gauge--fixing condition for the constraint~(\ref{P-}) is the light--cone
gauge~(\ref{lc-ga-1}).
As in the previous case we should accomplish the gauge--fixing procedure for this
gauge.
Note that the condition~(\ref{p-}) now follows from~(\ref{P+})--(\ref{P-tr}).
As a result, in physical variables the system~(\ref{act-mast}) is described by the
action
\begin{equation}\label{act-2-ph}
S_{mast}^{ph}=\int d\tau \left( P_+ \dot x^+ + \bar P^{(tr)} \dot x^{(tr)} +
P^{(tr)} \dot {\bar x}^{(tr)} + i(\dot{\bar\eta}\eta -\bar\eta\dot\eta) -
{\textstyle\frac{1}{P_+}}|P^{(tr)}|^2 \right).
\end{equation}
As follows from~(\ref{P+}), in this case ${P_+}>0\,$. Thus in the sector with
${P_+}>0$ the actions~(\ref{act-1-ph}) and~(\ref{act-2-ph}) are identical to each
other.


\begin{thebibliography}{96}
\bibitem{FrVas}
E.S.~Fradkin and M.A.~Vasiliev, Ann. of Phys. {\bf 177} (1987) 63; Phys. Lett. {\bf
B~189} (1987) 89.
\bibitem{Vas1}
M.A.~Vasiliev, Fortschr. Phys. {\bf 36} (1988) 33; Phys. Lett. {\bf B~243} (1990) 378,
{\bf B~257} (1991) 111.
\bibitem{FrLin}
E.S.~Fradkin and V.Ya.~Linetsky, Ann. of Phys. {\bf 198} (1990) 252.
\bibitem{SeS}
E.~Sezgin and P.~Sundell, JHEP {\bf 0109} (2001) 036
[{\tt hep-th/0105001}].
\bibitem{Vas}
M.A.~Vasiliev, Phys. Rev. {\bf D~66} (2002) 0660066 [{\tt hep-th/0106149}].
\bibitem{Vas-03}
M.A.~Vasiliev, {\it Higher-Spin Theories and $Sp(2M)$ Invariant Space--Time},
the talk presented at the third Sakharov Conference
in Physics, Moscow, 26-29 June 2002 [{\tt hep-th/0301235}].
\bibitem{Vas-a}
M.A.~Vasiliev, JHEP {\bf 0412} (2004) 046 [{\tt hep-th/0404124}].
\bibitem{Sor}
D.~Sorokin, {\it Introduction to the Classical Theory of
Higher Spins}, XIX Max Born Symposium ``Fundamental
Interactions and Twistor-like Methods'', Wroclaw, Poland,
28 Sept.-1 Oct. 2004, AIP Conf.Proc. {\bf 767}(2005)172
[{\tt hep-th/0405069}].
\bibitem{BekVas}
X.~Bekaert, S.~Cnockaert, C.~Iazeolla and M.A.~Vasiliev, {\it
Nonlinear Higher Spin Theories in Various Dimensions},
Lectures given at Workshop on Higher Spin Gauge Theories, Brussels, Belgium, 12-14 May 2004
[{\tt hep-th/0503128}].
\bibitem{BandLuk}
I.~Bandos and J.~Lukierski, Mod. Phys. Lett. {\bf A~14} (1999) 1257
[{\tt hep-th/9811022}].
\bibitem{BandLukSor}
I.~Bandos, J.~Lukierski and D.~Sorokin, Phys. Rev. {\bf
D~61} (2000) 045002 [{\tt hep-th/9904109}].
\bibitem{B-V}
I.~A.~Bandos, J.~A.~de Azc\'arraga, J.~M.~Izquierdo and J.~Lukierski,
Phys.\ Rev.\ Lett.\  {\bf 86} (2001) 4451 [{\tt hep-th/0101113}];\\
I.~A.~Bandos, J.~A.~de Azc\'arraga, J.~M.~Izquierdo, M.~Picon and
O.~Varela, Phys.\ Rev.\  {\bf D69} (2004) 105010
[{\tt hep-th/0312266}].
\bibitem{PST1}
M.~Plyushchay, D.~Sorokin and M.~Tsulaia, JHEP {\bf
0304} (2003) 013 [{\tt hep-th/0301067}].
\bibitem{BPST}
I.~Bandos, P.~Pasti, D.~Sorokin and M.~Tonin, JHEP
{\bf 0411} (2004) 023
[{\tt hep-th/0407180}].
\bibitem{BBAST}
I.~Ban\-dos, X.~Be\-ka\-ert, J.A.~de~Az\-c\'ar\-ra\-ga,
D.~So\-ro\-kin and M.~Tsu\-laia, JHEP {\bf 0505} (2005) 031
[{\tt hep-th/0501113}].
\bibitem{GSV}
O.A.~Gelfond, E.D.~Skvortsov and M.A.~Vasiliev, {\it Higher
Spin Conformal Currents in Minkowski Space}, {\tt
hep-th/0601106}.
\bibitem{FedLuk}
S.~Fedoruk and J.~Lukierski, Phys. Lett. {\bf B~632} (2006) 371
[{\tt hep-th/0506086}]; {\it Higher Spin Particles with Bosonic Counterpart of
Supersymmetry},
the Proceedings of the Int. Workshop SQS'05, Dubna, July 27-31, 2005, {\tt
hep-th/0512183}.
\bibitem{FedZim}
S.~Fedoruk and V.G.~Zima, JETP Lett. {\bf 61} (1995) 251; Class.
Quant. Grav. {\bf 16} (1999) 3653 [{\tt hep-th/9807192}]; J. Kharkov Univ.,
{\bf 585} (2003) 39 [{\tt hep-th/0308154}].
\bibitem{Fed}
S.~Fedoruk, {\it Twistor Transform for Spinning Particle},
XIX Max Born Symposium ``Fundamental Interactions and
Twistor-like Methods'', Wroclaw, Poland, 28 Sept.-1 Oct.
2004, AIP Conference Proceedings, {\bf 767}(2005)106.
\bibitem{Lecht}
C.~Devchand and O.~Lechtenfeld, Nucl. Phys. {\bf B~516} (1998) 255 [{\tt
hep-th/9712043}].
\bibitem{GuBl}
R.~Casalbuoni, Nuovo Cimento {\bf A~33} (1976) 389; \\
J.A.~de Azc\'arraga and J.~Lukierski, Z. Phys {\bf C~30} (1986) 221;
Phys. Rev. {\bf D~38} (1988) 509.
\bibitem{Fer}
A.Ferber, Nucl. Phys. {\bf B~132} (1978) 55.
\bibitem{ES}
Y.~Eisenberg and S.~Solomon, Nucl. Phys. {\bf B~309} (1988) 709; \\
Y.~Eisenberg, Phys. Lett. {\bf B~225} (1989) 95.
\bibitem{BC}
N.~Bengtsson and M.~Cederwall, Nucl. Phys. {\bf B~302} (1988) 81.
\bibitem{Town}
P.K.~Townsend, Phys. Lett. {\bf B~261} (1991) 65.
\bibitem{FIsoon}
S.~Fedoruk, E.~Ivanov, work in preparation.
\bibitem{GIOS}
A.~Galperin, E.~Ivanov, V.~Ogievetsky and E.~Sokatchev,
{\it Harmonic Superspace}, Cambridge University Press,
Cambridge, 2001, 306 p.
\bibitem{GIKOS}
A.~Galperin, E.~Ivanov, S.~Kalitzin, V.~Ogievetsky and E.~Sokatchev,
Class. Quant. Grav. {\bf 1} (1984) 469.
\bibitem{PenMac}
R.~Penrose and M.A.H.~MacCallum, Phys. Reports {\bf 6C} (1972) 241.
\bibitem{T}
P.K.~Townsend, Class. Quant. Grav. {\bf
8} (1991) 1231.
\bibitem{B}
I.~Bars, {\it Lectures on Twistors}, {\tt hep-th/0601091}.
\bibitem{SS}
E.~Sezgin and E.~Sokatchev,  Phys. Lett. {\bf B~227} (1989) 103.
\bibitem{BBS}
E.~Bergshoeff, M.~Blencowe and K.~Stelle,  Commun. Math. Phys. {\bf 128} (1990) 213.
\bibitem{HT}
P.S.~Howe and P.K.~Townsend, Class. Quant. Grav. {\bf
7} (1990) 1655.
\bibitem{IMT}
E.~Ivanov, L.~Mezincescu and P.K.~Townsend, {\it Fuzzy $CP(n|m)$ as a quantum
superspace},
In "Symmetries in Gravity and Field Theory", eds. V.~Aldaya, J.M.~Cerver\'o and
P.~Garci\'a,
Ediciones Universidad de Salamanca, 2004, pp. 385-408, {\tt hep-th/0311159}.
\bibitem{OS}
V.~Ogievetsky and E.~Sokatchev, Phys. Lett. {\bf B~79} (1978) 222.
\bibitem{IvLuk}
E.~Ivanov and J.~Lukierski, Phys. Lett. {\bf B~624} (2005) 304
[{\tt hep-th/0505216}].
\end{thebibliography}
\end{document}